# Glow discharge induced reactions in mixtures of ozone and chlorodifluoromethane with atmospheric gases

A. Dorn,[1] H. Mutaf,[2] R. Orhan,[2] W. Wolff,[3] T. Pfeifer,[1] H. Ahmedov [2]

[1]Max Planck Institute for Nuclear Physics, 69117 Heidelberg, Germany
[2]TÜBİTAK National Metrology Institute (UME), Kocaeli, Gebze, Türkiye
[3]Physics Institute, Federal University of Rio de Janeiro, Rio de Janeiro, Brasil

**Abstract**. The influence of extraterrestrial particles like cosmic radiation (CR) on the chemistry and ozone density in the Earth's stratosphere is not well investigated and normally neglected in stratospheric chemistry models. Here we present the commissioning of a lab-based apparatus which aims at simulating conditions in the stratosphere in order to get better insight into the reactions induced by the secondary-particle showers from high-energetic CR which can reach low altitudes. Admixtures of ozone and the halocarbon $CHClF_2$ (R22, chlorodifluoromethane) to atmospheric gases ($N_2$, $O_2$, Ar) were exposed to a glow discharge in the total pressure regime of a few hPa. According to the mass spectrometric analysis of the gas composition the discharge initiates significant ozone depletion by a factor four in the absence of R22. This depletion is strongly enhanced to two orders of magnitude in the presence of R22. The possible underlying reactions are discussed.

**Introduction**
There is a long history of research on the processes forming and modifying the ozone layer in the Earth stratosphere and on the ozone depleting influence of man-made halocarbons. The main driving force of the stratospheric ozone chemistry is the solar UV radiation. This radiation penetrates the higher-lying atmospheric layers (mesosphere, thermosphere) and is absorbed by oxygen $O_2$ and ozone $O_3$ in the stratosphere.

One question that is presently under debate is the influence of the energetic particle streams which originate from the Sun or from sources outside the solar system on the troposphere and biosphere. There is the solar particle stream (solar wind) from protons and alpha particles reaching energies up to roughly 500 MeV. Further, there is cosmic radiation (CR) which can have higher energies but has lower flux. Most primary and secondary particles are stopped in the ionosphere, which concerns essentially all solar particles. Only particle showers initiated by the more energetic CR (> 1 GeV) can reach lower atmospheric layers like the stratosphere. Here ionization rates are up to 100 $s^{-1}\,cm^{-3}$ and the ion density is around a few thousand $cm^{-3}$. There are observations that CR influences atmospheric properties such as cloud formation and ozone in the stratosphere [1]. In particular Liu demonstrated that stratospheric ozone concentration varies with the 11-year periodicity of the CR flux [2]. The significance of these observations and the mechanisms are under debate. Liu suggested a reaction model where anthropogenic chlorofluorocarbons (CFC) reaching the stratosphere are efficiently dissociated in reactions with secondary electrons from CR which are trapped on surfaces of stratospheric ice crystals [3].

On the modelling side there are simulations of the atmospheric ionization density as, e.g., the AtRIS code [4] which is a GEANT4 based simulation of CR induced atmospheric particle showers providing ionization densities as function of the incoming radiation energy and the altitude. This data can be the basis for modelling the atmospheric ion chemistry and the influence of CR showers on the relevant quantities like the ozone column density.

Works on ionization of atmospheric gas mixtures are scarce. Cacace et al. [5] have performed mass-spectroscopic ionization studies on mixtures of ozone and hydrochlorofluorocarbons (HCFCs) diluted in atmospheric gases. They found particular compound ions from ozone and HCFC which will be discussed below in relation with the present findings.

The abundant secondary electrons produced by the CR particle showers can be of importance for atmospheric ion chemistry. We aim at providing an experimental means to study the effects of electrons on ozone and the influence of HCFCs under conditions similar to the stratosphere. Therefore, we have set-up a laboratory apparatus consisting of a vacuum chamber that can be filled with controlled mixtures of gases at pressures between 1 and 4 HPa. A glow discharge is used to provide an electron gas to induce electron attachment, excitation and ionization processes and, thereby, initiate chemical processes modifying the gas concentrations. The evolving gas composition is analyzed by means of mass-spectroscopy. First studies were performed with the atmospheric gases ($N_2$, $O_2$) and admixtures of ozone and chlorodifluoromethane ($CHClF_2$) which is the main CFC component in the stratosphere [6]. We observe significant modifications of the gas concentrations and in particular a strong reduction of ozone.

**Experimental**

A scheme of the experimental setup is shown in Fig. 1. The experiments are carried out in a vacuum chamber with 0.4 m diameter and 0.5 m height with around 63 l volume. The gases are admitted with individual flux regulators allowing the adjustment of constant gas flows in a range from 10 to 1000 sccm (standard cubic centimeter per minute, 100 sccm = 1.689 hPa l/s). An ozone generator is utilized to produce around 2-3 % ozone admixture to the oxygen flow. During experiments the chamber pressure could be varied between 0.1 mbar and 4 mbar by adjusting the gas pumping speed. At in the lower pressure range the chamber is pumped by a turbo-molecular pump with an upstream adjustable throttle valve. In the higher pressure range the chamber is pumped directly by an oil-free scroll pump. To achieve the typical experimental conditions with 2 mbar chamber pressure and gas flows of 100-200 sccm a pumping speed of 0.8 – 1.6 l/s is required. Therefore, it takes around 40 – 80 seconds for a complete turnover of the chamber volume of 63 l.

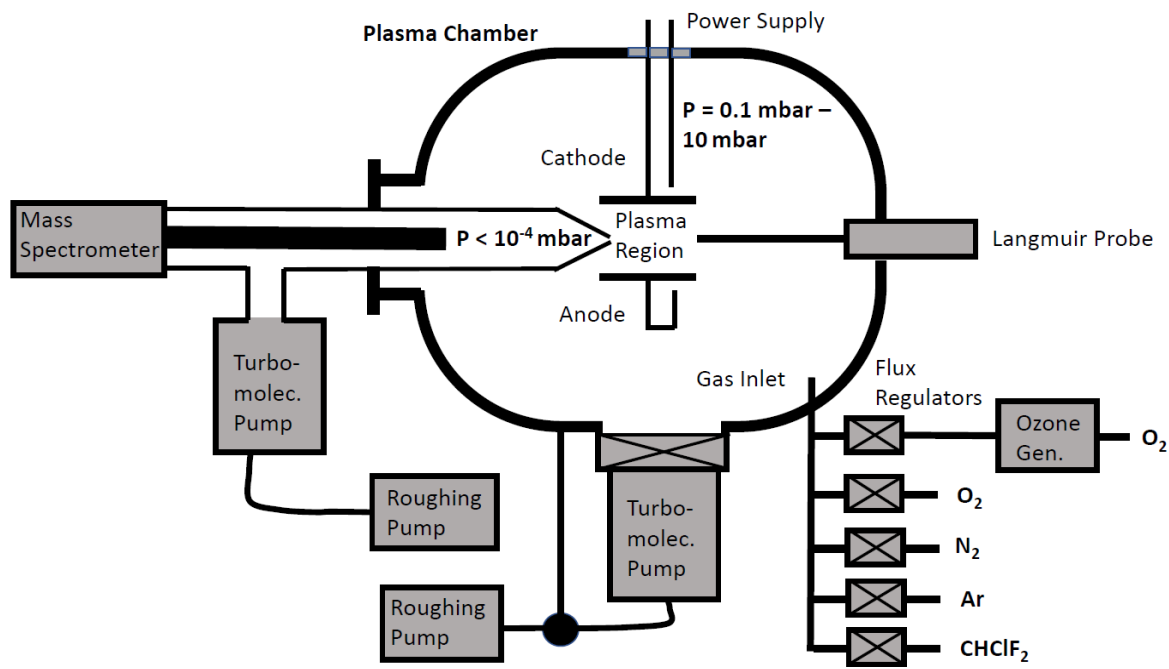

**Fig. 1:** Scheme of the experimental setup. For details see text.

To induce electron-molecule collisions a glow discharge is initiated between a cathode and an anode, both with 5 cm diameter and being 7 cm apart. Typical electrical voltages and currents at 2 mbar pressure are 700 V and 10 mA. A Langmuir probe and the entrance aperture of a mass spectrometer are positioned about halfway between cathode and anode. The commercial Langmuir probe (Impedans plasma measurement) is used to analyze the electron gas parameters which showed typical electron temperatures between 5 eV and 10 eV and electron densities in the range $(5 – 10) \cdot 10^7$ cm$^{-3}$. The gas composition at the edge of the plasma region is analyzed with a differentially pumped quadrupole mass spectrometer (Pfeiffer QMA200) with a 60 eV electron beam ion source. It is positioned in a chamber with a 30 cm long nozzle from PTFE material that can be moved close to or into the plasma region and allows gas to enter the MS chamber through an aperture of about 0.3 mm. Therefore, the MS probes mostly the neutral molecules which diffuse through the aperture into the differentially pumped MS where they get ionized and might fragment. Faraday cup ion current measurement was used. For identifying the individual molecular species, the ion masses listed in Table 1 were used. E.g. R22 ($CHClF_2$) shows various ionic fragments in the mass spectrum with the most abundant one being $CHF_2^+$ with about 90 % branching ratio. For identification of ozone ($O_3^+$, M = 48 amu) it must be accounted for that one R22-fragment with relative abundance of 0.7 % of the $CHF_2^+$-intensity has mass 48 amu.

Additionally, an optical spectrometer is attached to the chamber to analyze the plasma emission lines. It was not used in this study.

| Gas species | Ion species used for identification | M (amu) | Fraction of fragment channel | Other origin of ions |
|---|---|---|---|---|
| $N_2$ | $N_2^+$ | 28 | 91 % | - |
| $O_2$ | $O_2^+$ | 32 | 91 % | $O_3$ |
| $O_3$ | $O_3^+$ | 48 | 50 % | $CHClF_2$ (0.7 %) |
| $CHClF_2$ (R22) | $CHF_2^+$ | 51 | 90 % | $^{35}ClO$ |
| Ar | $Ar^+$ | 40 | 95 % | - |
| $^{37}ClO$ | $^{37}ClO^+$ | 53 | - | - |

**Table 1:** Atomic and molecular species for which the variation in time was studied by the MS (column 1). The second column shows the ion species used for identification.

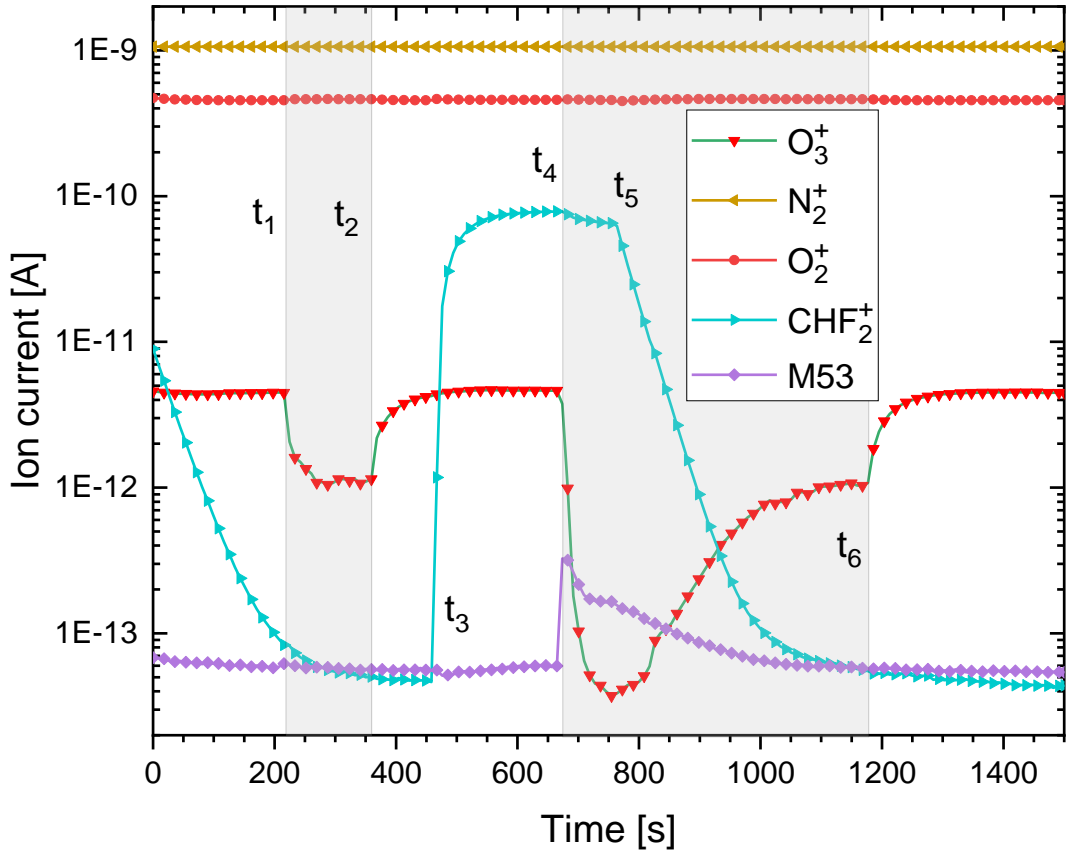

**Fig. 2:** Experimental ion current variations as function of time for gas fluences of 200 sccm $N_2$ (M = 28 amu), 100 sccm $O_2$ (M = 32 amu), 3 sccm $O_3$ (M = 48 amu). At times $t_1$ - $t_6$ the parameters are changed as follows: $t_1$, $t_4$: discharge on. $t_2$, $t_6$: discharge off. $t_3$: R22 flux on (11 sccm). $t_5$: R22 flux off. The total pressure is 1.8 mbar. Ion currents are normalized to the $N_2^+$ current at t = 0.

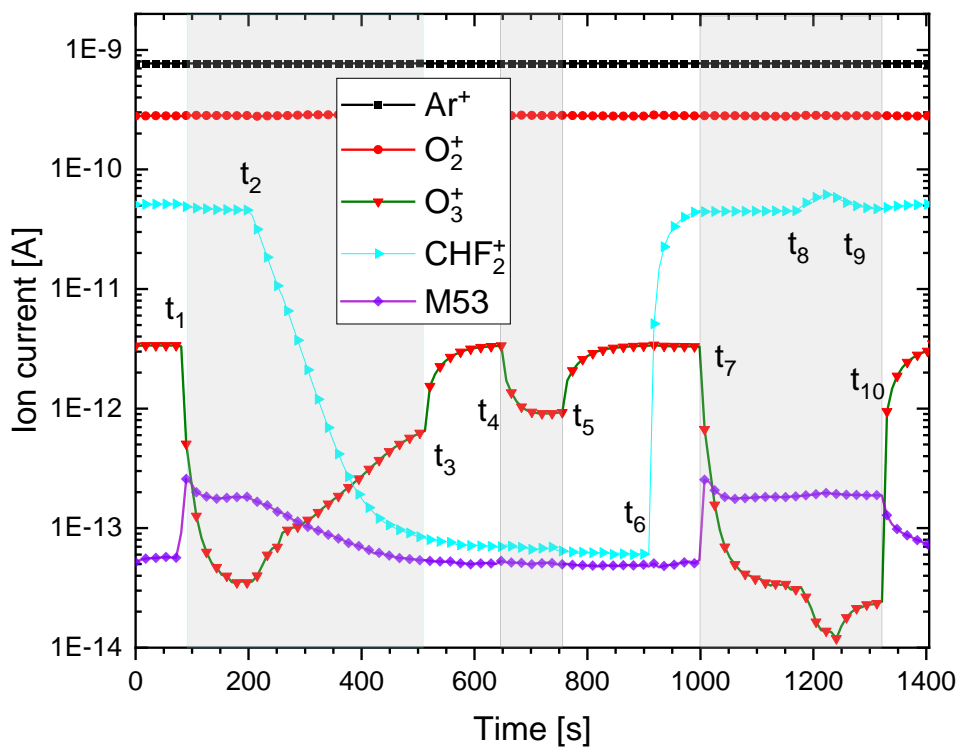

**Fig. 3:** Experimental ion current variations as function of time with gas fluences of 200 sccm Ar, 100 sccm $O_2$, 3 sccm $O_3$, 11 sccm R22. At times $t_1$ - $t_{10}$ the parameters are changed as follows: $t_1$, $t_4$, $t_7$ discharge on. $t_3$, $t_5$, $t_{10}$ discharge off. $t_2$: R22 flux off. $t_6$: R22 flux on. Between $t_8$ and $t_9$ the R22 flux was slightly increased. The total chamber pressure is 3 mbar. Ion currents are normalized to the $Ar^+$ current at t = 0.

**Results.**

In this study mixtures of atmospheric gases including ozone are exposed to electron collisions in a glow discharge. Small contributions of hydrochlorofluorocarbon R22 ($CHClF_2$) are introduced which is known to be harmful for ozone in the stratosphere. The MS ion currents are recorded for glow discharge on and off and with R22 flux on and off. Typical mass spectra covering all masses between 5 amu and 89 amu can be found in the Supplementary Information (SI). Here we focus on the behavior of the main gas components. In Fig. 2 an exemplary experimental sequence is presented. The ion yield currents taken with the MS are shown as a function of time. The initial gas flows were $N_2$: 200 sccm, $O_2$ 100 sccm. The $O_3$ flow was roughly 2 sccm which was deduced from the relative MS ion currents for $O_3^+$ and $O_2^+$ and the partial ionization cross sections to produce $O_2^+$ from $O_2$, as well as $O_3^+$ from $O_3$. This number is a lower limit since part of the ozone can be lost on the way to the MS by conversion to $O_2$ on chamber wall surfaces. At $t_1$ the glow discharge is started (voltage 900 V, current 10 mA). The electron gas temperature was around 6.5 eV and the electron density about $5 \cdot 10^7$ cm$^{-3}$. A significant reduction of $O_3$ concentration by a factor of four is observed. In the same time a small increase of the $O_2^+$ signal of 1.6 % is found. After switching off the discharge at $t_2$ the $O_3^+$ ion current recovers to the original value. Subsequently at $t_3$ the gas CHClF$_2$ is admitted with a low flux of 11 sccm which is 3.7 % of the combined N$_2$ and O$_2$ fluxes. In the MS this gas is identified with the ionic fragment CHF$_2^+$ at m = 51 amu which is the most abundant ion for ionization of R22 (90 %). At $t_4$ the glow discharge is switched on and this results in a strong decrease of ozone by more than two orders of magnitude. Again, there is a small increase of the $O_2^+$ signal by 1.1 %. In the same time also the CHClF$_2$ concentration is reduced by about 15 %. Then at $t_5$ the R22 gas flux is stopped allowing the ozone ion yield to recover to the original value for plasma operation. Finally, when the plasma is stopped at $t_6$ the ozone concentration settles at the starting value. Additionally, an important observation is the course of the ion with M = 53 amu. In the stratosphere the relevant mechanism for ozone depletion by CFCs is the release of chlorine initiating the odd chlorine catalytic cycle which involves ClO as the first product (see discussion below). Therefore, we examined mass 53 corresponding to $^{37}$ClO (25 % abundance) while $^{35}$ClO (75 % abundance) is hidden under the dominant peak with M = 51 from CHF$_2^+$. Indeed, there is significant increase in 53 amu yield under plasma operation if both $O_3$ and R22 flux are on. This leads us to the tentative conclusion that M = 53 amu is a reaction product of both molecules where at least one fragment from an electron induced dissociation is involved.

In order to test if reactions involving nitrogen play a role in another experiment N$_2$ was replaced by argon and a similar sequence was performed as shown in Fig. 3. Initially Ar, O$_2$ and R22 gases were on and at $t_1$ the discharge was started giving rise to a decrease of ozone by two orders of magnitude. At $t_2$ R22 was shut off resulting in ozone increase. At $t_4$ and $t_5$ the discharge is switched on and off again, showing an ozone decrease by about a factor of four. The dependence of the ozone concentration on R22 flux is demonstrated during discharge operation $t_7$ - $t_{10}$ where R22 is briefly increased between $t_8$ and $t_9$. This results in a further decline of ozone.

These sequences demonstrate the strong influence of electron collision induced processes on the ozone concentration. Further reaction sequences with various gas combinations, gas fluxes and at different pressure regimes are presented in the SI.

## Discussion

There are different chemical processes known to evolve in atmospheric gas mixtures with ozone-depleting substances under the influence of ultraviolet radiation or under charged particle impact as in the present case. The basic reaction cycle responsible for ozone production and the absorption of UV-light is the Chapman cycle:

Ozone production: $O_2 + h\nu$ (< 242.4 nm) $\rightarrow 2O$ $\Rightarrow$ $O + O_2 \xrightarrow{M} O_3$ (1)

Ozone destruction $O_3 + h\nu$ (< 320 nm) $\rightarrow O + O_2$ $\Rightarrow$ $O_3 + O \rightarrow 2\, O_2$ (2)

In principle both reaction sequences can also be induced in the glow discharge where electron collisions can initiate the molecular excitation steps in (1) and (2). Reactions (2) are the most likely processes for ozone depletion in our study if R22 is absent. This is indicated by the $O_2^+$ ion current increase, which is consistent with the respective ozone decrease within the accuracy of the measurement. While in the stratosphere there is an equilibrium of $O_3$ production and $O_3$ destruction we observe significant stronger discharge-induced destruction. This is most likely due to the present pressure conditions in the hPa range which is about a factor of ten lower than in the stratosphere. Consequently, the second step in reaction (1) where a third particle M is required to stabilize $O_3$ has a much lower reaction rate than in the stratosphere. In addition to the dissociative excitation of reaction (2) other possible electron induced reactions with ozone are ionization (E > 12.53 eV) and ionization with subsequent dissociation (E > 12.8 eV) producing $O_2^+$ or $O^+$ [7]. Finally, the more abundant low energy electrons in the glow discharge can undergo dissociative electron attachment (DEA) with ozone between 0 and 3 eV energy yielding $O_2^-$ and $O^-$ [8]. The negative ions can neutralize in collisions with positive ions or surfaces and an effective ozone depletion takes place.

If R22 is introduced a number of new fragments are produced from collisions involving free electrons. The possible reactions are dissociative electron attachment (DEA) [9], dissociative excitation [10] and dissociative ionization [11]. In all these processes the most likely product is the chlorine atom or the negative chlorine ion together with the molecular fragment $CHF_2^{(+)}$. For ionization we have measured the absolute partial ionization cross sections for various CFCs including R22 from low to high impact energies [12]. The dominant fragmentation results in $CHF_2^+$ + Cl which is just the channel used for detection of R22 according to table 1. Chlorine in the stratosphere is very efficient in ozone depletion which evolves in catalytic cycles. In the odd chlorine cycle the following reactions take place

$$Cl + O_3 \rightarrow ClO + O_2 \qquad (3)$$

followed by

$$ClO + O \rightarrow O_2 + Cl. \qquad (4)$$

In our experiment there are signatures that at least the first step of this cycle is playing a significant roll due to the detection of $^{37}ClO$ in the mass spectrum as mentioned above. The second step depends on the concentration of O which can be produced in electron collisions with $O_3$ and $O_2$ analogous to eqns. (1) and (2).

Another ozone-depleting reaction in the stratosphere is the odd nitrogen cycle involving NO and $NO_2$:

$$NO + O_3 \rightarrow NO_2 + O_2 \qquad (5)$$

$$O + NO_2 \rightarrow NO + O_2 \qquad (6)$$

In the stratosphere the nitrogen oxides mainly originate from nitrous oxide $N_2O$ coming from the troposphere but partially also from CR which ionizes and dissociates $N_2$ into radicals which further react with $O_2$ [13]. In principle ionization/dissociation of $N_2$ can also be a source of nitrogen oxides in our experiment. Therefore, we have replaced the $N_2$ gas by argon and an exemplary experimental sequence is shown in Fig. 3. We found within the accuracy of the MS data the same behavior of the ozone concentration as with nitrogen gas. For discharge operation and with R22 gas supply the ozone reduction is two orders of magnitude while for absence of R22 the ozone concentration is reduced by a factor four. Therefore, the nitrogen oxides do not seem to play a significant role in the present experiment. One reason might be the large $N_2$ binding energy of 9.7 eV in combination with the relatively low temperature of the electron gas which results in a low $N_2$ dissociation rate.

Another ozone depleting mechanism was found in a previous study by Cacace et al. [5] using advanced analytical and mass spectroscopic methods. The authors have found reactions between ionized ozone and CFCs which in case of R22 are:

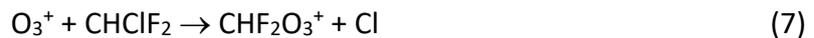
$$O_3^+ + CHClF_2 \rightarrow CHF_2O_3^+ + Cl \qquad (7)$$

Here the ion is unstable and its main fragmentation channel is producing neutral CO.

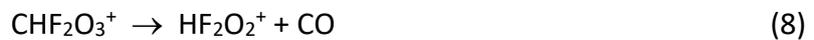
$$CHF_2O_3^+ \rightarrow HF_2O_2^+ + CO \qquad (8)$$

CO with M = 28 has the same mass as $N_2$ such that we cannot separate it with the present MS. Nevertheless, we have performed mass scans under various conditions as they are shown in the Supplementary Information Figs 1-7. For a mixture of 400 sccm $O_2$ and about 12 sccm $O_3$ and 11 sccm R22 we find that the M = 28 amu peak intensity, which is due to residual gas, increases for discharge operation by around 50 % with respect to the discharge being switched off. The M = 28 amu peak is unchanged with/without discharge if R22 is not introduced. Therefore, a certain quantity of CO is produced from reactions like eqns. (7) and (8). In the overview mass spectra we found indications of more reaction products from ozone and R22 which are initiated by the discharge like $ClFO^+$ (M = 70 amu, 72 amu) and $F_2O^+$ (M = 33 amu). This supports the findings of Cacace at al. that there is a rich chemistry between ions and neutrals of oxygen, ozone and F22.

In order to conclude on processes in the Earth stratosphere several aspects have to be considered. In the stratosphere the ionization density by CR is much lower than in the glow discharge, namely around 10-100 $cm^{-3} s^{-1}$ and the resulting electron-ion pair density is around $10^3 – 10^4$ $cm^{-3}$. In comparison the electron-ion density in the present experiment is above $10^7$ $cm^{-3}$. Therefore, the present experiment strongly enhances the collisional reactions in the gas mixture. On the other hand, the catalytic ozone depleting cycles are running on longer time scales and cannot contribute significantly to modify the ozone concentrations due to the fast gas exchange rate in the present experiment. From these observations possible future studies should aim at higher chamber pressure of 10 to 20 hPa and lower gas turnover rates to better mimic stratospheric conditions. As result of these conditions significantly lower halocarbon concentration can influence ozone concentration.

## Summary

In conclusion we studied the influence of a glow discharge on mixtures of atmospheric gases ($O_2$, $N_2$) containing ozone and a typical halomethane R22 ($CHClF_2$). The pressure range was a 1 – 3 hPa in order to simulate conditions in the upper stratosphere. The gas composition was analyzed with mass spectroscopy and the electron-gas temperature was in the range of 5 – 10 eV with densities 5 – 10 ·$10^7$ $cm^{-3}$. In absence of R22 the discharge initiates ozone depletion by factor of four which is enhanced to two orders of magnitude by addition of about 3 % of R22 flux. The possible underlying collisional and chemical processes were discussed. The reactions involving nitrogen do not seem to play a significant role.

Possible future studies should analyze the pressure dependence of the observations. E.g. at higher pressure the discharge-induced ozone depletion can be compensated by ozone production in three-body processes O + $O_2$ + M → $O_3$ + M. Also, the gas exchange rate can be reduced in order to enhance the role of the repeating catalytic cycles with respect to the initial collisional processes. In this way the natural stratospheric conditions can be approached and extrapolations to lower halomethane concentrations and lower electron densities can be performed.

In addition, modelling of the processes starting from the various collisional reactions initiated by low energetic electrons would be desirable. In particular existing published simulations of the atmospheric ionization density apply simplified assumptions, e.g., applying a single average energy loss - the W-value - for formation of an electron-ion pair [4]. In order to facilitate setting up simulations with a more accurate description we provide an open source data repository with cross sections for electron impact ionization and electron attachment for the various atmospheric gases including chlorofluorocarbons [14].

**Appendix**

We present mass spectra which cover a large mass range from 4 to 89 amu which were taken at different conditions as given below.

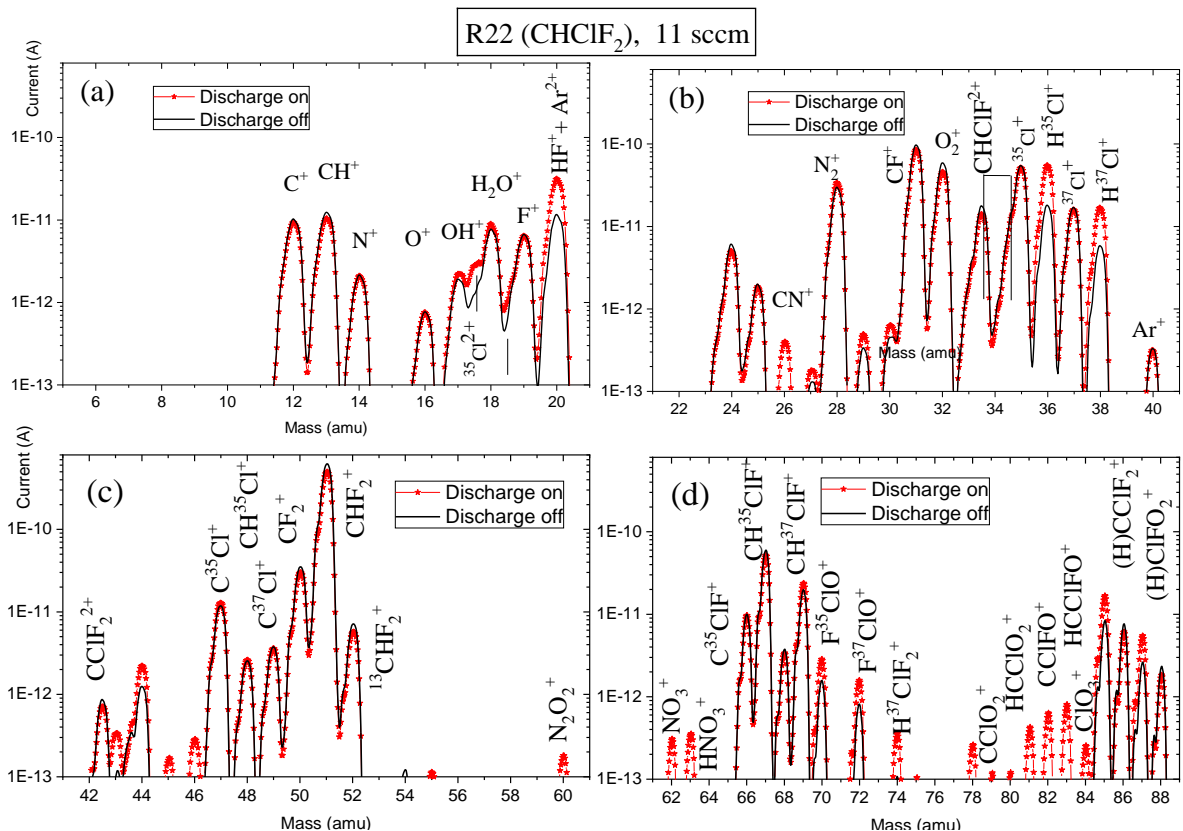

**Fig A1:** Mass spectrum for the flux of 11 sccm of R22 corresponding to 0.05 HPa. Measurement with glow discharge off (black solid line) and with glow discharge on (red symbols).

Fig. A1 shows the mass spectrum for R22 gas only in the plasma chamber with and without the glow discharge. For ionization in the mass spectrometer ion source R22 produces a range of ionic fragments from which M = 51 amu $CHF_2^+$ is the most intense. Therefore, this fragment was used to monitor the time evolution of R22. Interesting is to see which intensities change for discharge operation. $CHF_2^+$ is slightly decreasing due to reduction of the R22 concentration from ionization and dissociation processes. The main resulting product atomic chlorine is not reaching the mass spectrometer but HCl is increasing significantly. The same is observed for F which is observed as HF (M = 20 amu). On the other hand, the mass over charge ratio 17.5 amu and 18.5 amu are increasing corresponding to doubly charged $^{35,37}Cl^{2+}$. At mass 48 amu there is a R22 fragment which has to be accounted for when analyzing the ozone concentrations.

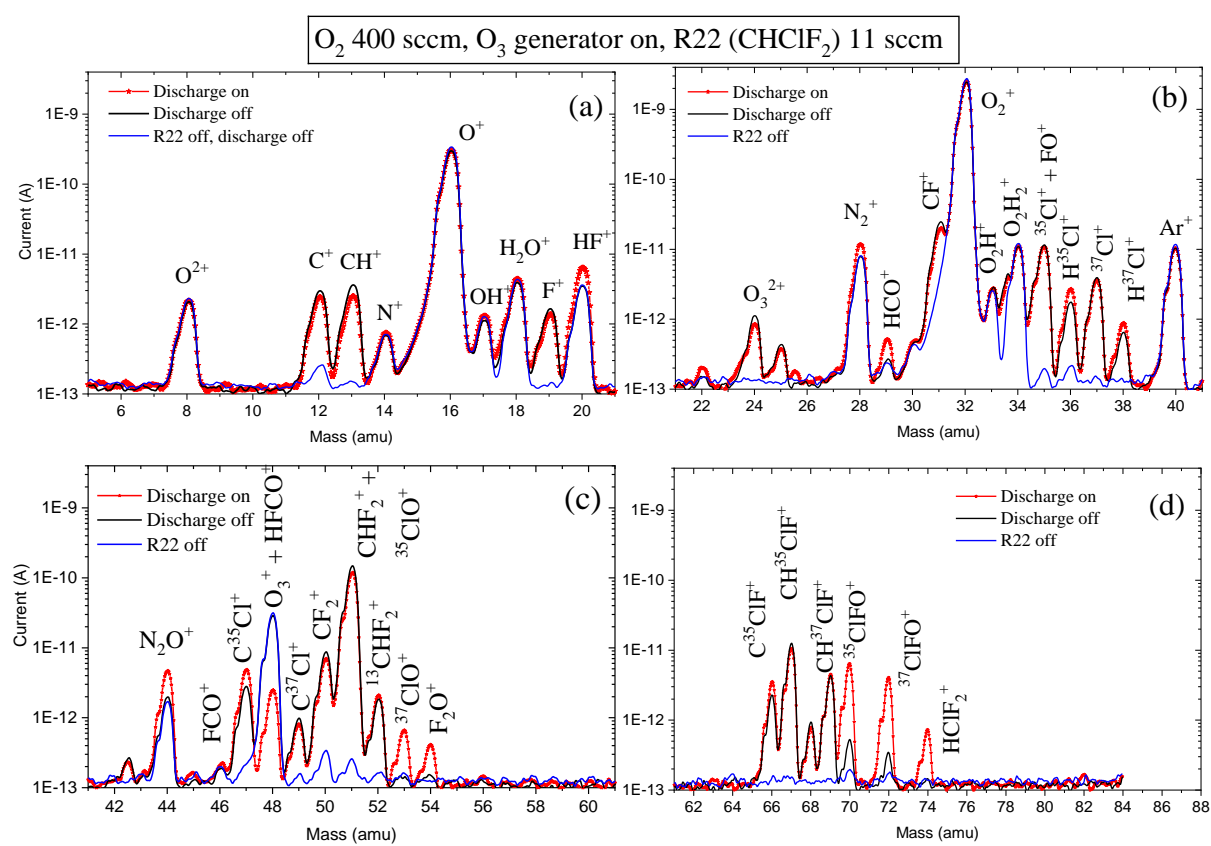

**Fig A2:** Mass spectrum for the 400 sccm flux of $O_2$, about 7 sccm $O_3$ and 11 sccm of R22 at the total pressure 2.4 HPa. Measurement with glow discharge off (black solid line) and with glow discharge on (red symbols). Furthermore, a scan with R22 flux off and discharge off is shown (blue solid line).

Fig. A2 shows mass spectra for oxygen with admixtures of ozone and R22. Again the changes due to the discharge can be recognized.

## Data availability

The data supporting this article have been included as part of the Supplementary Information.

## Acknowledgements

The project 21GRD02 BIOSPHERE has received funding from the European Partnership on Metrology, co-financed by the Horizon Europe Research and Innovation Program of the European Union and the participating States. Funder ID: 10.13039/100019599. Grant number: 21GRD02 BIOSPHERE.

## References

[1]     H. Svensmark and E. Friis-Christensen, J. Atmos. Terr. Phys. 59, 1225 (1997); H. Svensmark, Phys. Rev. Lett. 81, 5027 (1998); N. D. Marsh and H. Svensmark, ibid. 85, 5004 (2000).

[2]     Q. B. Liu and L. Sanche, Phys. Rev. Lett. 87, 2001

[3]     Qing-Bin Lu; *AIP Advances*  **11,** 115307 (2021)

[4]     Banjac, S., Herbst, K., & Heber, B. (2019), Journal of Geophysical Research: Space Physics, 124(1), 50.

[5]     F. Cacace, G. de Petris, F. Pepi, M. Rosi, A. Troiani, Chem. Eur. J. (2000)**, 6**, 2572; F. Cacace, G. de Petris, F. Pepi, M. Rosi, A. Sgamellotti, Angw. Chem. Int. Ed. **38**, 2408 (1999).

[6]     S. A. Montzka, B. D. Hall, J. W. Elkins, Geophys. Res. Lett. **36**, L03804 (2009)

[7]     Newson, Int J Mass Spec Ion Processes, 148 (1995) 203

[8]     G. Senn, J. D. Skalny, A. Stamatovic, N. J. Mason, P. Scheier, and T. D. Märk, Phys. Rev. Lett. **82**, 5028

[9]     P. Cicman, A. Pelc, W. Sailer, S. Matejcik, P. Scheier, T.D. Märk, Chemical Physics Letters 371 (2003) 231–237

[10]    J. F. Ying; K. T. Leung, J. Chem. Phys. 105, 2188–2198 (1996), https://doi.org/10.1063/1.472092

[11]    L Sigaud N. Ferreira, L H Coutinho, V L B de Jesus and E C Montenegro  2012 J. Phys. B: At. Mol. Opt. Phys. 45 215203

[12]    M. Dogan, W. Wolff, D. M. Mootheril, T. Pfeifer and A. Dorn, Phys. Chem. Chem. Phys. DOI: 10.1039/d5cp00746a; W. Wolff, M. Dogan, H. Luna, L. H. Coutinho, D. Mootheril, W. Baek, T. Pfeifer and A. Dorn, Rev. Sci. Instrum., 2024, 95, 095103.

[13]    M. Nicolet, Planet. Space Sci. 1975. Vol. 23, p. 637

[14]    Dorn, A., Upendranath, P. (2025). https://doi.org/10.5281/zenodo.15398189